\documentclass{elsart}
\usepackage{natbib}
\usepackage[dvips]{graphicx}
\usepackage{longtable}

\begin{document}
\begin{frontmatter}

\title{Direct measurements of neutron capture on radioactive isotopes}

\author{A.~Couture},
\ead{acouture@lanl.gov}
\author{R.~Reifarth} 
\ead{reifarth@lanl.gov}
\address{Los Alamos National Laboratory, Los Alamos, NM, 87545, USA}
\maketitle
\begin{abstract}
We simulated the response of a 4$\pi$ calorimetric $\gamma$-detector 
array to decays of radioactive isotopes
on the $s$-process path. The GEANT~3.21 simulation package was used. 
The main table contains estimates on the maximum sample size
and required neutron flux based on the latest available neutron capture cross section at 
30~keV. The results are intended to be used to estimate the feasibility of 
neutron capture measurements with 4$\pi$ arrays using the time of flight technique.
\end{abstract}


\end{frontmatter}

\tableofcontents{}

\section{Introduction}
The scope of this work is to provide a tool for experimentalists to 
quickly estimate the feasibility of a direct neutron capture (n,$\gamma$) 
measurement using a calorimeter to detect the emitted 
$\gamma$-rays. Such measurements are typically performed with the 
time-of-flight (TOF) technique, which allows the determination of the 
cross sections as a function of incident
neutron energy. Particularly,  the need for data in the keV neutron 
energy region where the competing elastic 
scattering channel is very strong drove the trend to detect the emitted 
prompt $\gamma$-rays after the capture event with 4$\pi$ calorimeters \cite{WGK90a,RBA04,CAA06}. 
The Q-value for capture events is unique for each isotope, but typically ranges from 
4-10~MeV.  The decay multiplicity following capture typically ranges from three to six,
though there are some nuclei where low multiplicity (one or two) decay is preferred.  
By exploiting knowledge of the capture signature, 
scattering as well as radioactive decay events can then very efficiently be discriminated 
from neutron capture events based on the total energy deposited in the detector
and the observed multiplicity of the events.
The TOF technique offers advantages over activation techniques in that it is not limited
to isotopes with a long-lived product. Furthermore, as the data needs change over time,
the cross-section for the desired neutron distribution can be determined whereas
activation methods determine a cross-section for the specific distribution used in the
experiment

Calorimetric detectors have close to 100\% total detection efficiency for $\gamma$-rays with energies 
between $\approx$100~keV and a few~MeV. The decay of the radioactive samples inside a calorimeter
implies a certain count rate, which depends on the detection efficiency and the decay properties
of the isotopes under investigation. Experimental techniques such as passive shielding 
and increased energy-thresholds on the individual detectors can be used to
lower the rate from the intrinsic activity of the source.

The isotopes under investigation in this paper are of interest for the $s$-process nucleosynthesis. 
About 50\% of the element abundances beyond iron are produced via slow neutron capture 
nucleosynthesis ($s$ process). Starting at iron-peak seed, the $s$-process mass flow 
follows the neutron rich side of the valley of stability. 
%
%
%
The isotopes capture neutrons and become more neutron rich until the neutron capture
lifetime exceeds the beta-decay lifetime, at which point the isotope will $\beta^-$-decay 
to a more 
stable neighbor which has a higher Z and is less neutron rich.  This stair-step 
procedure allows the $s$ process to work to higher and higher atomic and 
mass numbers.  The reader is referred to \cite{Kae99} for a review on the $s$ process.  
The position of this stair-step edge depends upon the actual
neutron density in the stellar scenario.  For particular nuclei, the beta-decay rate
and the neutron capture rate are close enough that there is a branching,
allowing some of the material to capture an additional neutron while the rest of
it decays before an additional capture can take place.  These so-called 
``branch-point'' isotopes are particularly interesting since they can provide the
tools to constrain modern stellar $s$-process models.  Experimental neutron
capture cross-sections of all of the isotopes on the $s$-process paths are needed,
but the cross-sections of the branch-point isotopes are particularly useful in
constraining the temperatures and neutron densities at the nucleosynthesis
site.  Unfortunately, the radioactive nature of the branch-point isotopes has
made them difficult to study in the laboratory.    

\begin{figure}
  \includegraphics[width=.75\textwidth]{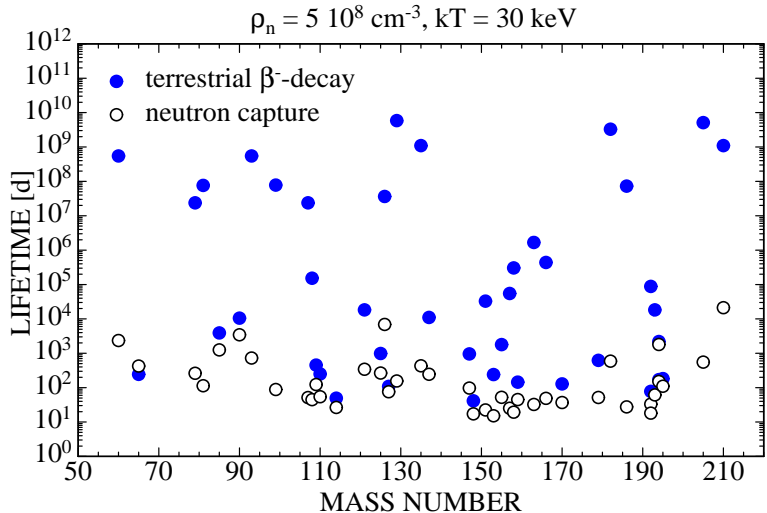}
  \caption{Neutron capture life times (open circles) and terrestrial $\beta^-$ life times 
  (filled circles) for unstable isotopes on the classical $s$-process path as 
  a function of mass number. Shown are only isotopes where the neutron capture is faster than the
  stellar $\beta^-$ decay for a neutron density of 5$\cdot$10$^8$~cm$^{-3}$ at a temperature of 30~keV. 
  The neutron capture cross sections are taken from the references given in the table and the 
  stellar decay rates from \cite{TaY87}.
  \label{isotopes_classic}}
\end{figure}

In a recent estimate, the neutron density within the classical $s$-process model 
\cite{KBW89} was calculated to be $n_n~=~(4.94^{+0.60}_{-0.50})\times 10^8~{\rm cm^{-3}}$ 
\cite{RAH03}. Figure~\ref{isotopes_classic} shows a summary of the neutron capture and 
$\beta^-$ decay times for radioactive isotopes on the neutron rich side of the 
valley of stability, under the condition that the classical neutron capture occurs 
faster than the terrestrial $\beta^-$ decay.  The vast majority of isotopes 
where an experimental neutron capture cross section is desirable have 
$\beta^-$ half-lives of at least hundreds of days, though some isotopes of
interest have half-lives of as little as tens of days. 

\begin{figure}
  \includegraphics[width=.75\textwidth]{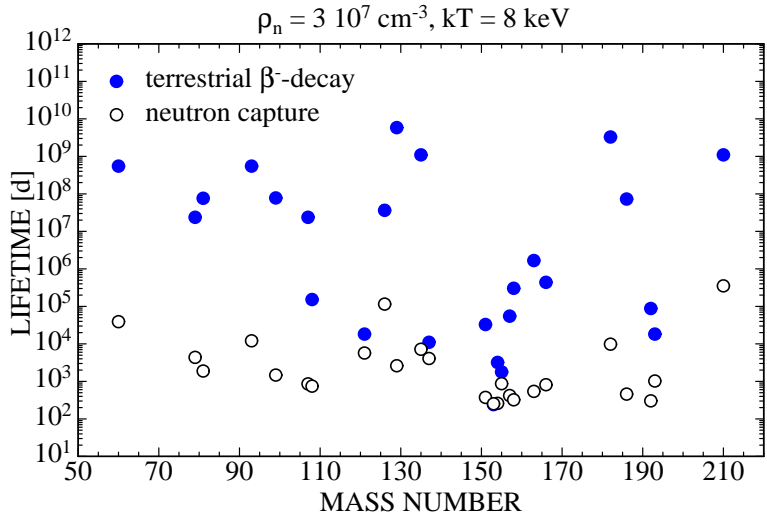}
  \caption{Neutron capture live times (open circles) and terrestrial $\beta^-$ live times 
  (filled circles) for unstable isotopes on the classical $s$-process path as 
  a function of mass number. Shown are only isotopes where the neutron capture is faster than the
  stellar $\beta^-$ decay for a neutron density of 3$\cdot$10$^7$~cm$^{-3}$ at a temperature of 8~keV. 
  The neutron capture cross sections are taken from the references given in the table and the 
  stellar decay rates from \cite{TaY87}.
  \label{isotopes_c13}}
\end{figure}

\begin{figure}
  \includegraphics[width=.75\textwidth]{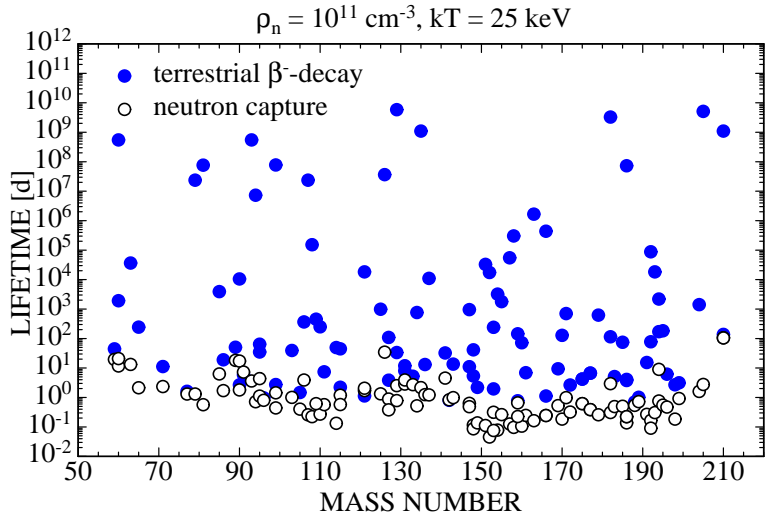}
  \caption{Neutron capture live times (open circles) and terrestrial $\beta^-$ live times 
  (filled circles) for unstable isotopes on the classical $s$-process path as 
  a function of mass number. Shown are only isotopes where the neutron capture is faster than the
  stellar $\beta^-$ decay for a neutron density of 10$^{11}$~cm$^{-3}$ at a temperature of 25~keV. 
  The neutron capture cross sections are taken from the references given in the table and the 
  stellar decay rates from \cite{TaY87}.
  \label{isotopes_ne22}}
\end{figure}

The modern picture of the main $s$-process component refers to the He shell burning phase in 
AGB stars \cite{LHL03}. The $s$ process in these stars experiences episodes of low neutron
densities of about $3\cdot10^7$~cm$^{-3}$, the $^{13}$C($\alpha$,n)-phase, and very high 
neutron densities, the $^{22}$Ne($\alpha$,n) phase.  
The highest neutron densities during the latter phase reach values of up to 10$^{11}$ cm$^{-3}$. 
Fig.~\ref{isotopes_c13} and Fig.~\ref{isotopes_ne22} show the same as Figure~\ref{isotopes_classic}, 
but for the conditions during the two phases of the main component of the $s$ process in AGB stars. 
During the $^{22}$Ne($\alpha$,n) phase, the lifetime versus neutron capture is much shorter
resulting in isotopes with half-lives of just a few days forming the critical branching
points for the $s$-process reaction flow.

Improved experimental techniques, especially as far as the neutron source and sample preparation are 
concerned, are necessary to perform direct neutron capture measurements on such 
isotopes \cite{RHH04, WHK06}. The Detector for Advanced Neutron Capture Experiments (DANCE) 
is currently one of the strongest combinations of $\gamma$-ray detector and neutron source \cite{RBA04}.
It is designed as a high efficiency, highly segmented 4$\pi$ BaF$_2$ 
detector for calorimetrically detecting 
$\gamma$-rays following a neutron capture. DANCE is 
located on the 20~m neutron flight path 14 (FP14) at the Manuel Lujan Jr. Neutron Scattering 
Center at the Los Alamos Neutron Science Center (LANSCE) \cite{LBR90}. 
Neutrons are produced via spallation
reactions induced by 800 MeV protons hitting a tungsten target. 
The design of the detector is such that a
full 4$\pi$ array would consist of 162 crystals of four different 
shapes, each shape covering the same solid
angle. The BaF$_2$ crystals form a  spherical shell with an inner radius of 
$\approx$17~cm and a thickness of 15~cm.
Two crystals are removed in order to allow the neutron
 beam to enter and exit, so that the 
array has 160 elements as used in a typical experiment.
The neutron flux per energy decade available at the current
position is approximately 3$\cdot$10$^5$~n/s/cm$^2$. 
Measurements can be performed between approximately 
10~meV and 200~keV \cite{RBA04}. 

The Total Absorption Calorimeter (TAC) at CERN can perform (n,$\gamma$)
experiments with samples sizes of 10-100~mg. 
TAC is also a 4$\pi$ BaF$_2$ detector array, but with 42
crystals to cover the entire solid angle (40 in use during the experiments). 
The BaF$_2$ crystals form a  spherical shell with an inner radius of 
$\approx$10~cm and a thickness of 15~cm. TAC is located at a 185~m flight path at nTOF. 
Neutrons are produced via spallation reactions induced by 20~GeV protons hitting 
a lead target \cite{BCD03, MAA06}. 

The newly established Frankfurt Neutron Source at the Stern-Gerlach-Zentrum (FRANZ) at the university of
Frankfurt, Germany, will provide pulsed keV-neutron fluxes in the order of 10$^6$~n/s/cm$^2$. A 4$\pi$
BaF$_2$ array for detecting the $\gamma$-rays following a neutron capture is planned.

These arrays all use flash ADCs with sample rates of 500-1000MHz to directly
digitize the response from the detectors.  This offers great flexibiliy in
event reconstruction in software offline.

This paper gives estimates of the maximum number of radioactive atoms which
could be tolerated by the DANCE array as well as the number of neutrons at 
the sample required for a direct measurement using a DANCE-like detector. 
While similar calculations could be done for any geometry, the authors had
greatest access to the DANCE array, and thus, could most easily verify that
the decay simulation produced realistic events with this detector.  The
basic assumption underlying all of the simulations and estimates is that
the rate of pile-up in the calorimeter and the individual detectors is 
sufficiently low as to be negligible.  While the pile-up effects could be 
calculated for a particular instrument under very high rate conditions for
a particular experiment, the results would not be very portable as they would
necessarily depend strongly on the details of how the acquisition is accomplished
as well as the number and layout of the individual crystals for a particular 
detector at a particular site.  Since the goal of this work was to provide
a reference that would allow a quick check on what would be required for
a measurement at a wide range of facilities, the decision was made to 
give values for which pile-up effects should not play a role.  
 
It should be
straight-forward for the
reader to scale the numbers given in the table to the specific conditions of other experiments.
Suggestions for how that scaling may be done and which factors may be important
can be found in the detailed discussion of the simulations for $^{191}$Os.
More details will be given in the next sections. The radio-isotopes under investigation are 
isotopes heavier than $^{56}$Fe and fulfill the condition 
that at least 10\% of the expected reaction flow is through neutron capture 
during the $^{22}$Ne phase.

\subsection{Monte Carlo Simulations} 
The results presented in the table are based on Monte Carlo simulations of 
the decay cascades and the DANCE detector array.

First, decay cascades according to the decay schemes of each isotope (\cite{Fir96}) 
were generated using a Monte Carlo method. The energy distribution of the 
emitted electrons or positrons 
was calculated by means of the Fermi theory of beta decay. 
Following \cite{Kra88}, the internal conversion coefficients
 were estimated according to 

$\alpha_{conversion}=\frac{l}{l+1} \alpha^4 Z^3 \left[\frac{2 m_e}{E_{\gamma}}\right]^{l+2.5}, 
\mbox{if}~E_{\gamma} > E_k \\ $
$\alpha_{conversion}=\frac{l}{l+1} \frac{\alpha^4 Z^3}{8} \left[\frac{2 m_e}{E_{\gamma}}\right]^{l+2.5}, 
\mbox{if}~E_{\gamma} \le E_k \\ $

as a function of the multipolarity ($l$), the fine structure constant ($\alpha$), 
the charge of the product nucleus ($Z$), the mass of the electron ($m_e$), the energy of the
transition ($E_\gamma$), and the ionization energy of the k-shell electron ($E_k$). 
The probability for electron conversion is then:

$p_{conversion}=\frac{\alpha_{conversion}}{1+\alpha_{conversion}}$

In order to simplify the simulations, an "average multipolarity" of $l=1.7$.
was assumed.  
Then these cascades were
started inside the DANCE ball and the response of the array was simulated using the GEANT~3.21 
detector simulation package (\cite{GEA93}).  Reference (\cite{RBA04}) can
be consulted for more details on the simulated geometry. 

Since a 6~cm $^6$LiH shell is usually used at DANCE to reduce the
background from scattered neutrons, the array was simulated with and without
the $^6$LiH shell. A similar sphere is in place inside TAC at n-TOF, but made of 
C$_{12}$H$_{20}$O$_4$($^6$Li)$_2$ (\cite{CAA06}). In addition, 
simulations were also done with one, two, and 
five millimeters of lead serving as a passive shield around the sample to reduce
the radio-decay background.  All of the simulations with lead shielding included the $^6$LiH shell.
A total of one million cascades were randomly created for each isotope. Each cascade was 
then simulated five times with randomly chosen angular distributions
under each of the five geometric conditions.
Individual detector thresholds were set in software in 100~keV intervals
to investigate the additional reduction of the count rate from the decay of the source.  

\begin{figure}
  \includegraphics[width=.75\textwidth]{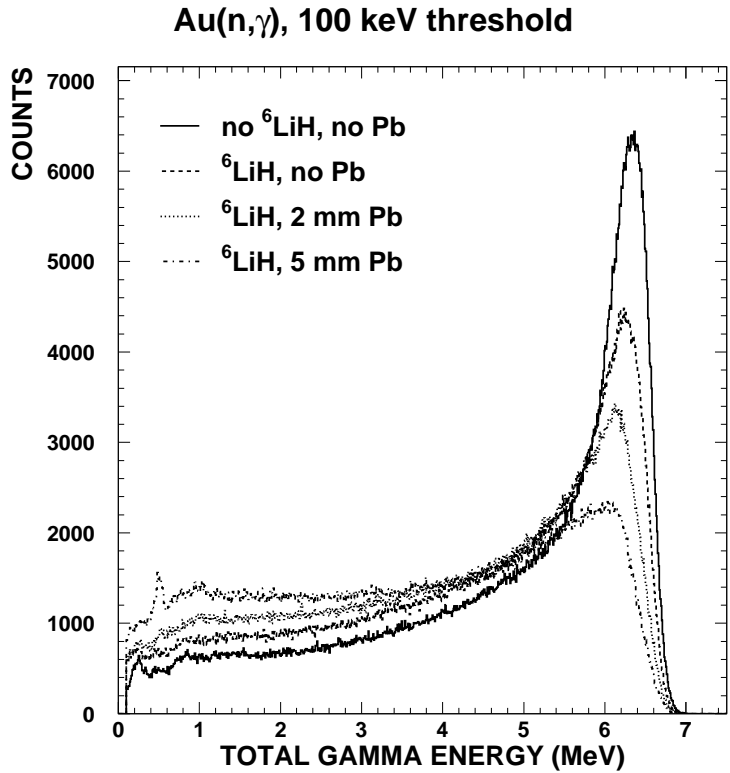}
  \caption{Sum energy spectrum for $^{197}$Au(n,$\gamma$) events for different absorbing materials and 
  100~keV single detector threshold. One million capture events were simulated. 
  \label{gold_absorbers}}
\end{figure}

\begin{figure}
  \includegraphics[width=.5\textwidth]{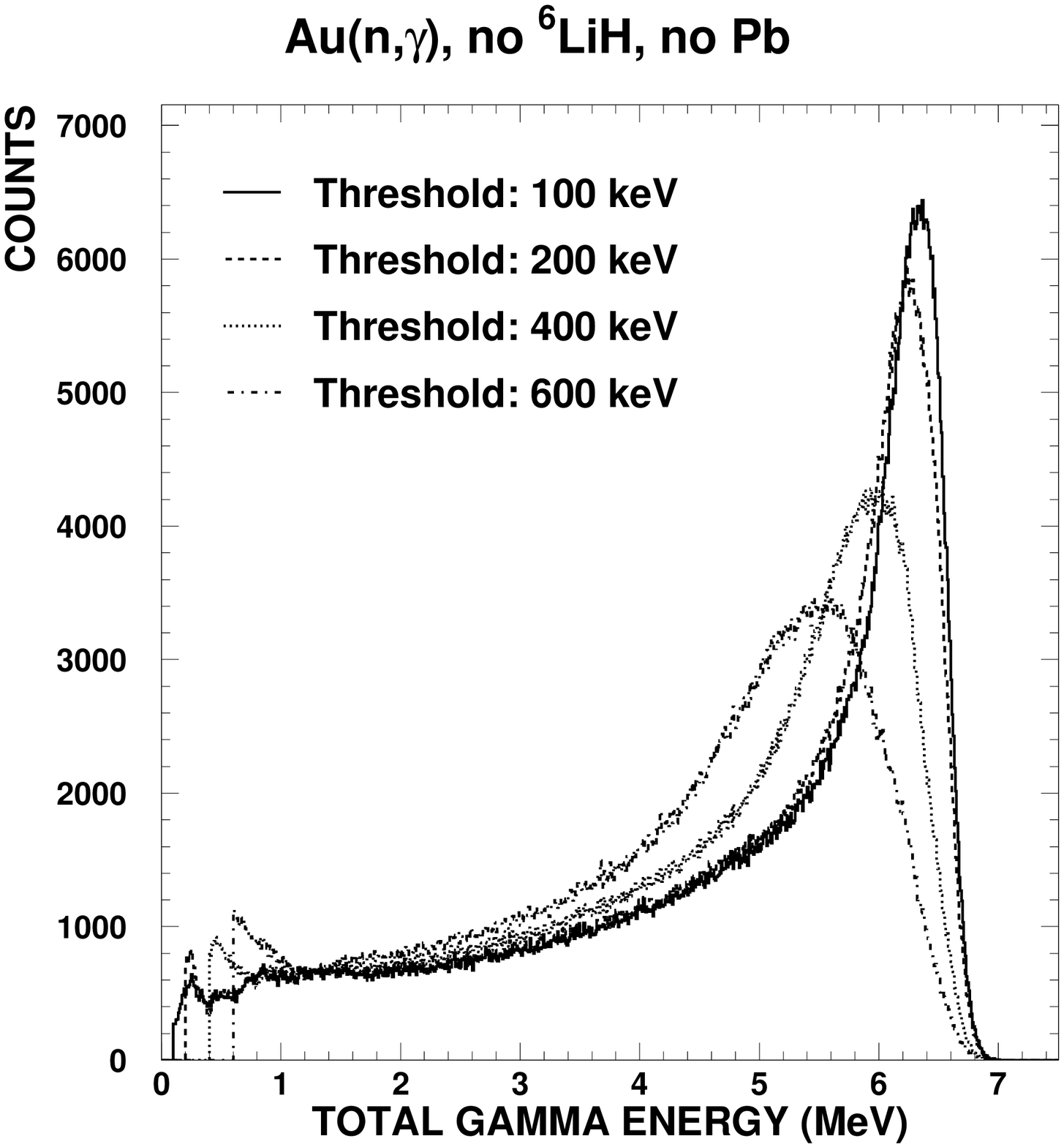}
  \includegraphics[width=.5\textwidth]{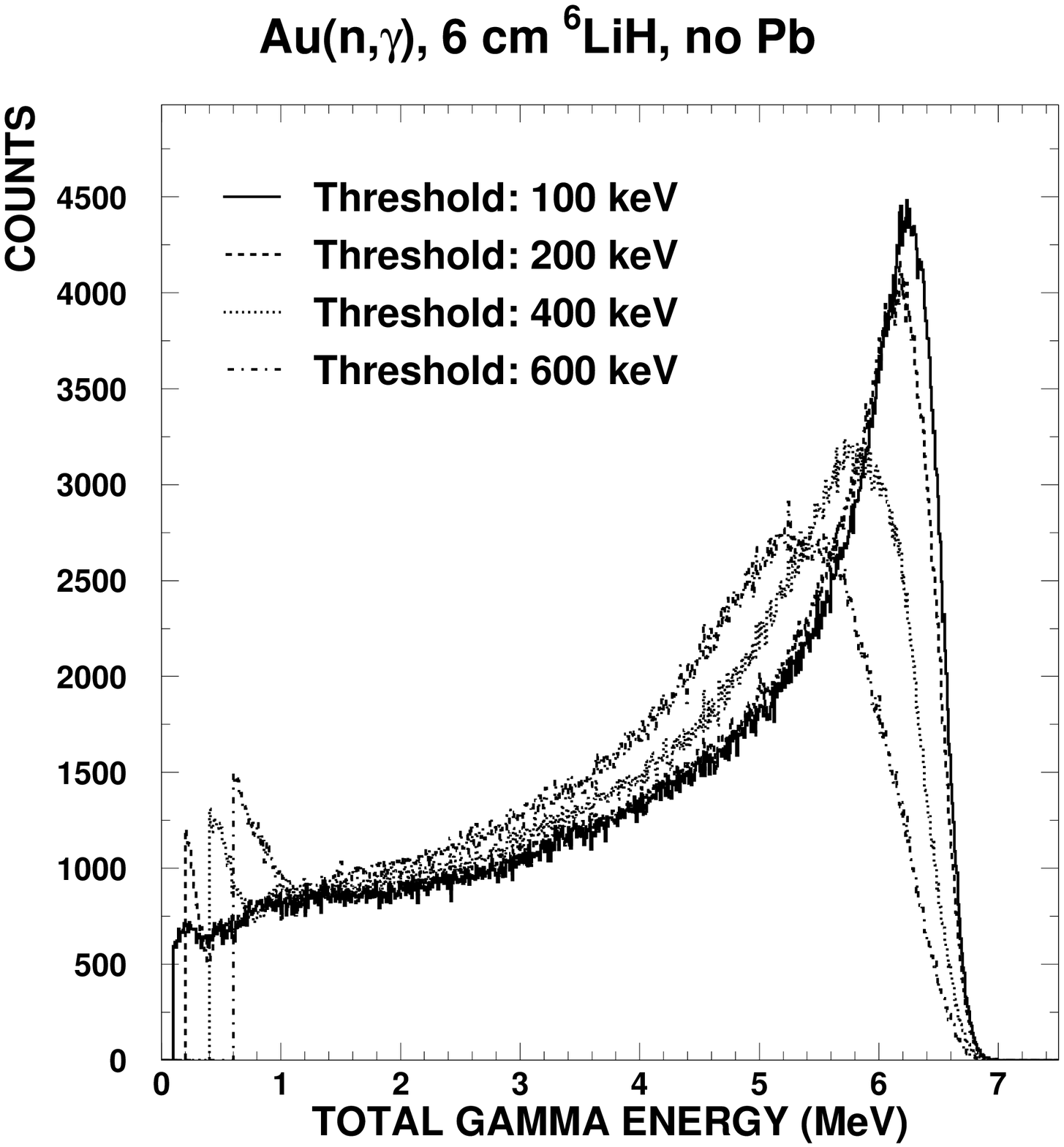}
  \includegraphics[width=.5\textwidth]{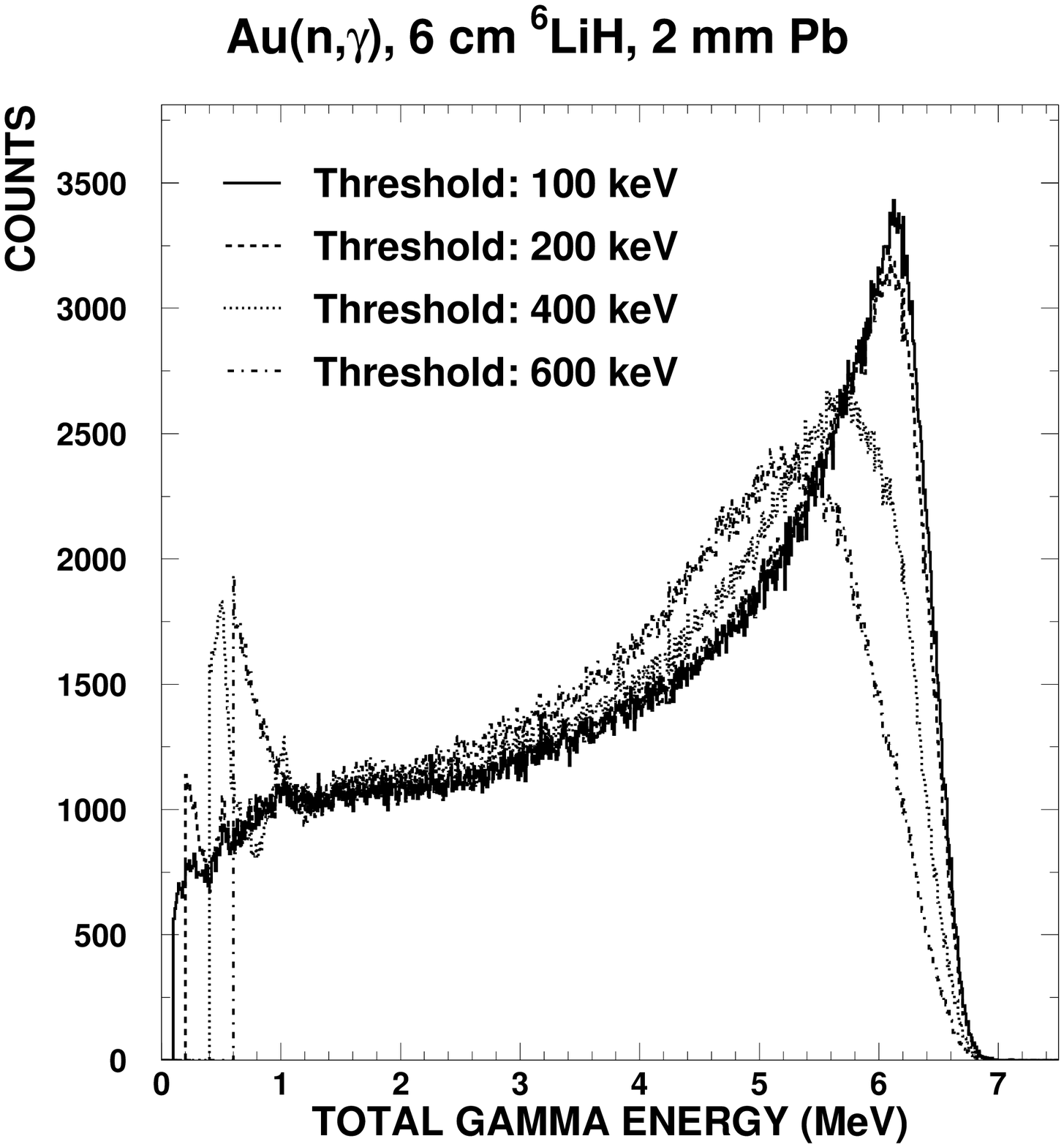}
  \includegraphics[width=.5\textwidth]{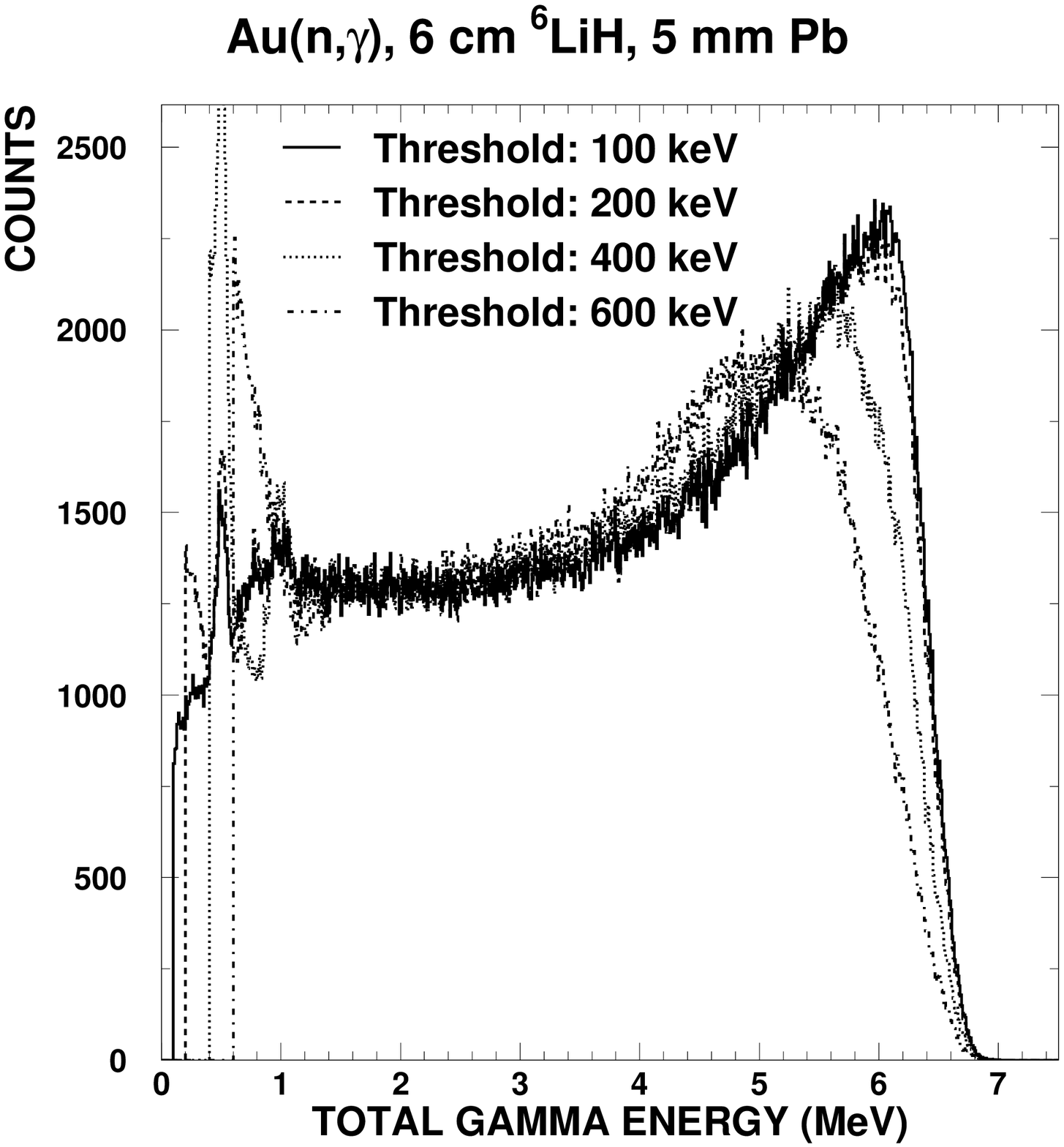}
  \caption{Sum energy spectrum for $^{197}$Au(n,$\gamma$) events for different absorbing materials and 
  variable single detector threshold. One million capture events were simulated for each setup.
  \label{gold_thresholds}}
\end{figure}

Absorbing material between sample and detectors does not only affect the radiation from the 
radioactive decay, but also $\gamma$-rays following a neutron capture. Since these are the desired 
signature for the capture event, better shielding or increased thresholds will always be a trade-off
between decreased sensitivity to background and decreased efficiency for capture events. 
Figure~\ref{gold_absorbers} shows the effect of different thicknesses of lead around the sample on the 
energy spectrum of the 4$\pi$ array following neutron captures on gold \cite{bec00} 
for a single-detector threshold of 100~keV.
Figure~\ref{gold_thresholds} shows the response of the array to gold captures for different absorbers
and different thresholds.
 
In many cases, the radioactive decays populate daughters which either have 
metastable states or which are themselves radioactive.  
If the lifetime of the daughter state was shorter than the one hundred days,
then the daughter and the parent were assumed to be in 
equilibrium.  This condition was considered a reasonable determination
on the ability to separate the decay products from the parent.
The decay rate of the daughter was equal to the rate of the 
decay which populated the daughter since the two were in equilibrium.
The final rate of observations in the array was the sum
of the rate of observations from the parent and the
observations from any daughters.

\subsection{Neutron capture cross sections}  

The neutron capture cross-sections used for determining which isotopes
were included in the compilation as well as estimating the
necessary neutron flux for an improved measurement was based on the 30~keV
Maxwellian averaged cross-section (MACS).  The cross-sections were based on the
most recent data available as of January 2006.  For most isotopes,
this meant that the values were taken from \cite{BBK00}.
There were several isotopes where more recent measurements have
been made, however.  These were included when available.  If values
were not available from \cite{BBK00} or a recent measurement, then
the values were calculated from \cite{JEFF30}.

\subsection*{Acknowledgments}
Many thanks to R.C.~Haight for his helpful comments on the manuscript and to J.M.~O'Donnell for his indispensable
help in enabling the use of lots of CPU time.
This work has benefited from the use of the Los Alamos Neutron Science Center at the Los Alamos 
National Laboratory. This facility is funded by the US Department of Energy and operated 
by the University of California under Contract W-7405-ENG-36. 
This work was funded by the LANL LDRD-ER project 20060357.
This project was supported through 
the Joint Institute of Nuclear Astrophysics by NSF-PFC grant PHY02-16783.

\section{Explanation of Tables} 
\subsection{General Remarks}
The first row of Table~\ref{isotopes} 
gives the isotope and the Maxwellian Averaged Cross Section (MACS)
at $kT~=~30$~keV including the reference for the cross section.

The numbers in columns 2-5 and rows 4-8 provide the maximum number of atoms 
tolerable inside a 4$\pi$ array (simulated for the example of DANCE) for each 
isotope under investigation ($N_{max}$). $N_{max}$ depends on the following 
three variables:

\begin{itemize}
  \item{$R_{max}$, the maximum rate tolerable in the array.  For the DANCE array, the primary
    limitation is the coincidence time window.
    $R_{max}$\ was chosen to be $R_{max}=10^7$~s$^{-1}$} for
    these calculations, consistent with the conditions mentioned in the Introduction
    regarding pile-up.  With the 160 detectors in DANCE, this would correspond to a 
    rate of $\approx 10^5$\ in each detector
    for low multiplicity decays.  For a less segmented detector,
    the primary limitation
    may come from the rate in the individual detectors rather than the rate in the array since
    the individual detector signals will have lengths of approximately 500~ns.
    More details can be found on determining this parameter for a particular setup in
    the detailed discussion of the $^{191}$Os case.
  \item{$q$, the quality factor of the array for the decay cascade, given by the 
    ratio of the number of observed cascades to the number of simulated cascades.
    The number of cascades observed is dependent on the condition that
    the energy deposition in a single detector is above a given threshold. This number comes 
    from the Monte Carlo simulations and depends strongly on the decay properties of the isotope.
    A quality factor of $q$=1.0 would indicate that every cascade was observed while a quality factor
    of 0.01 would indicate that 99\% of the cascades 
    deposited no energy above threshold in any detector.}
    A quality factor of $q>$1 is possible and would indicate that on average multiple detectors were
    above threshold for each cascade.
  \item{$\tau$, the life-time of the isotope versus radiodecay.  The lifetime is related to the
    half-life by $\tau$~=~t$_{1/2}$/ln(2).}
\end{itemize}

\begin{equation}
N_{max} = \frac{R_{max}\, \tau}{q}
\end{equation}

The columns refer to different single-detector thresholds in keV and the rows correspond to 
different absorbers between sample and detectors. 

The last column gives the required neutron flux in (neutrons/cm$^2$/s/energy-decade)
for each geometry and single-detector threshold 
of 100~keV for a desired capture rate of 0.1~s$^{-1}$/energy-decade. The unit
(neutrons/cm$^2$/s/energy-decade) is especially useful for neutron spectra at spallation 
sources (LANSCE, n-TOF), since the resulting spallation neutron spectrum has typically a 
high-energy part following a 1/E dependence, hence the number of neutrons per energy
decade will be constant.  
This would correspond to approximately 10,000 events/energy decade over the course
of a two week experiment. This is a typical number of statistics needed in order
to improve the status of the data.
For comparison, the neutron flux available at the sample position of DANCE is approximately 
3$\cdot$10$^5$~neutrons/cm$^2$/s/energy-decade, at the sample position 
of TAC at n-TOF approximately 10$^5$~neutrons/cm$^2$/s/energy-decade are available.

\subsection{Detailed Example: $^{191}$Os}
In order to further clarify the steps which were taken both in the simulations
and in the determination of the maximum number of atoms tolerable, the
calculation for $^{191}$Os will be done explicitly. $^{191}$Os beta-decays to a 
metastable state of $^{191}$Ir.  The half-life of $^{191}$Os is 15.4 days, while the
lifetime of the metastable state of $^{191}$Ir has a lifetime of 4.94 seconds \cite{Fir96}.
Since the lifetime of the daughter is so much shorter than the lifetime of the parent,
the ratio of $^{191}$Os:$^{191m}$Ir will quickly come in to equilibrium and the 
decay rate of $^{191m}$Ir will be approximately the decay rate of $^{191}$Os, though the two will
be asynchronous.  The decay rate $\lambda$\ is given by $\lambda$~=~ln(2)/t$_{1/2}$,
or $\lambda\, =\, 5.21 \times 10^{-7}$~decays/sec/atom. 

\begin{figure}
  \includegraphics[width=.75\textwidth]{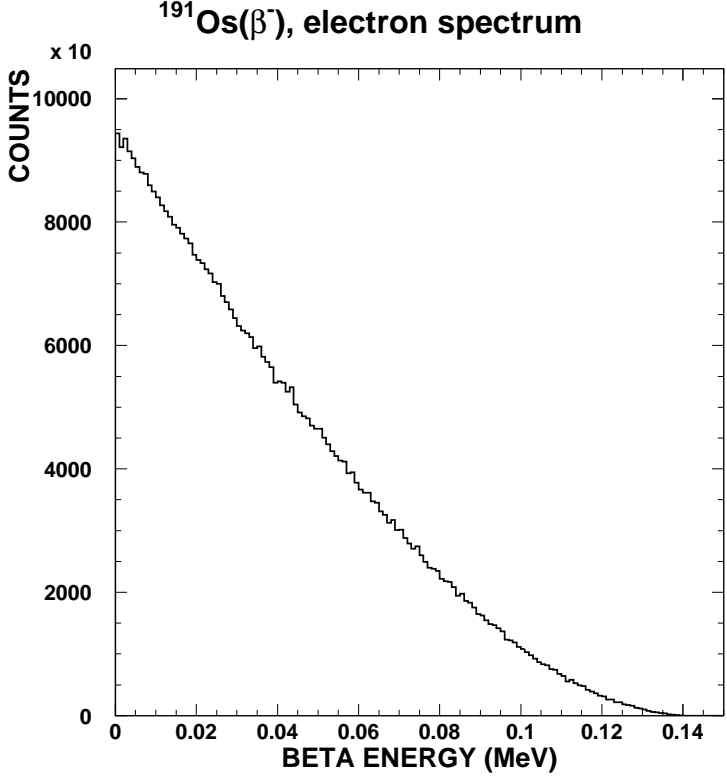}
  \caption{Electron spectrum for $^{191}$Os($\beta^-$) as used for the GEANT simulations. 
  \label{os_beta_decay}}
\end{figure}

\begin{figure}
  \includegraphics[width=.75\textwidth]{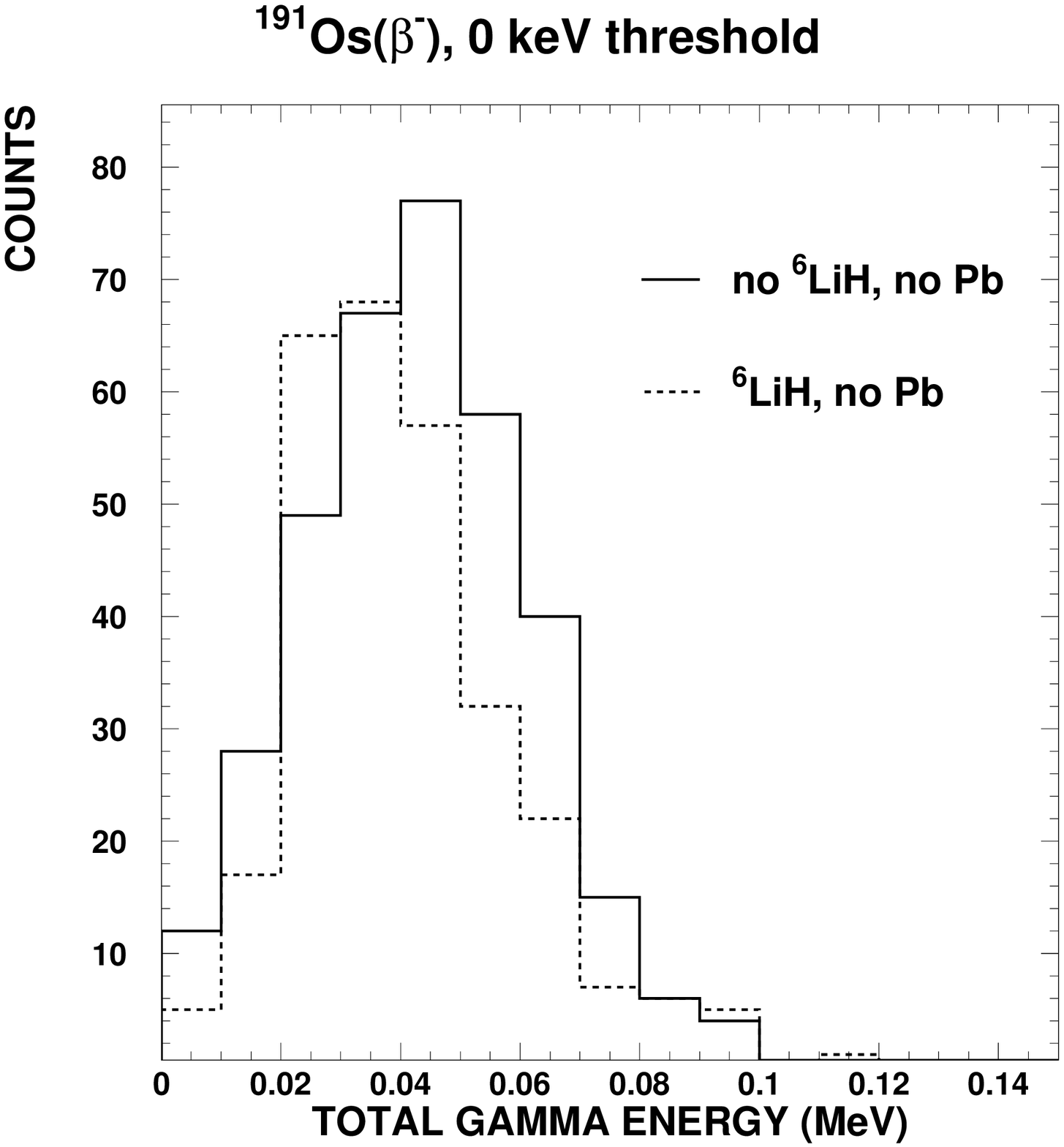}
  \caption{Response of the 4$\pi$ array to decays of $^{191}$Os($\beta^-$). A threshold of 100~keV
  discriminates almost all events. Five million decay events were simulated.
  \label{os_g_cuts}}
\end{figure}

The first step is then to generate decays for both the osmium and the iridium and 
determine the detector response to the decay.  The osmium will be discussed first
and then the iridium.  One million decays were generated in a 
Monte Carlo method for the beta decay of $^{191}$Os. The simulated beta-decay spectrum can
be seen in Fig.~\ref{os_beta_decay}. Each cascade was then started five times, meaning a total 
of five million events were simulated using GEANT. 
Because of the low kinetic energy of the beta,
even before placing cuts on the detectors, there is very little response from the detector
array.  Fig.~\ref{os_g_cuts} illustrates the array sum energy for a threshold of 0~keV and
different thicknesses of $^6$LiH. All other simulated combinations of geometry and threshold resulted 
in no detected events.
The detection probability was determined by calculating
the ratio of events seen in the array to decays simulated. Table~\ref{explicit}
follows this discussion and contains the information from all of the simulations. 

\begin{figure}
  \includegraphics[width=.75\textwidth]{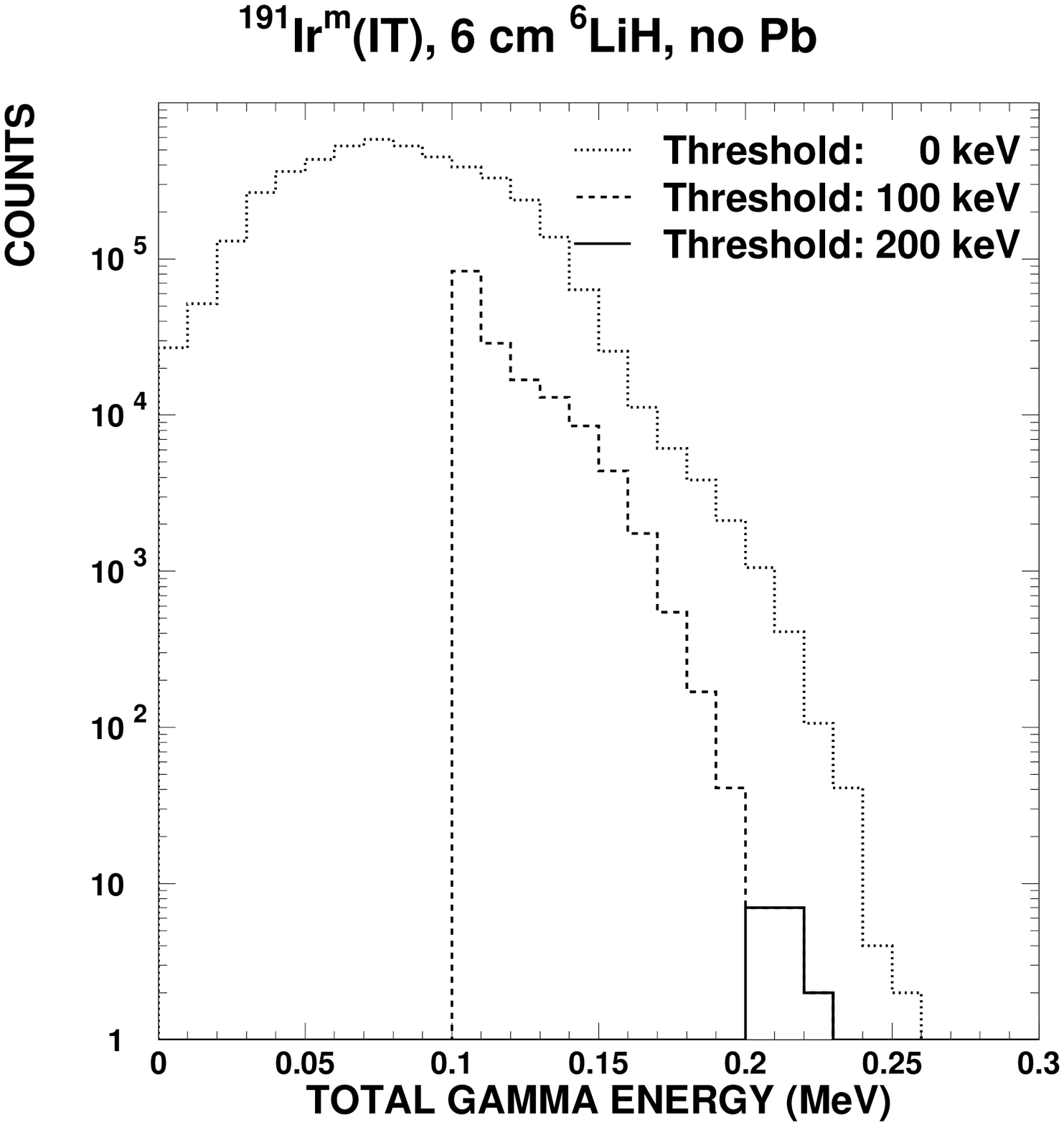}
  \caption{Response of the 4$\pi$ array to decays of $^{191}$Ir$^m(\beta^-)$ with different thresholds. 
  No lead, but 6~cm $^6$LiH were assumed between sample and detector. Five million decay events were simulated.
  \label{ir_t_cuts}}
\end{figure}

\begin{figure}
  \includegraphics[width=.75\textwidth]{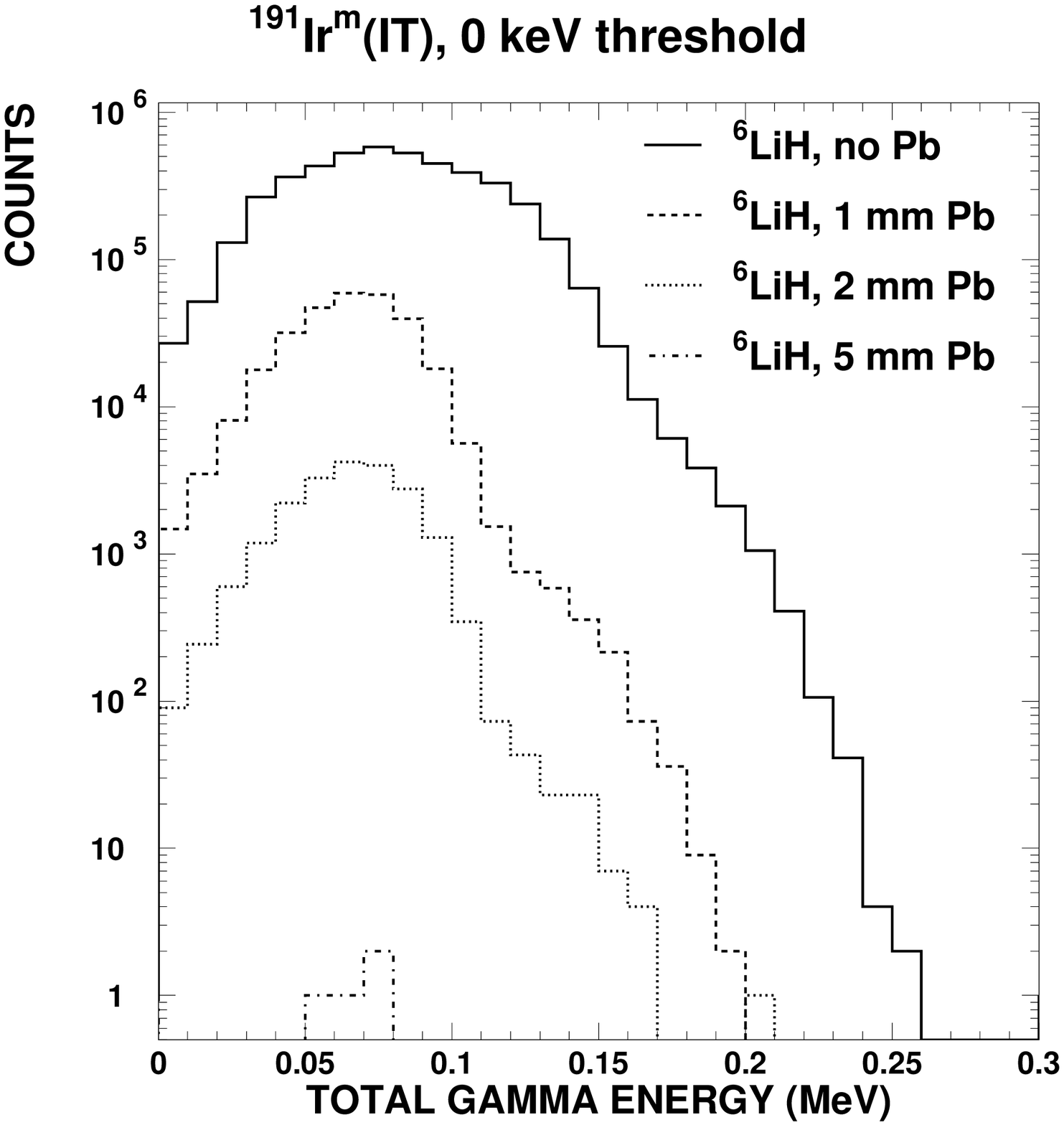}
  \caption{Response of the 4$\pi$ array to decays of $^{191}$Ir$^m(\beta^-)$. 
  A threshold of 0~keV and different geometries are shown. The case without $^6$LiH is 
  very similar to the one with $^6$LiH, no Pb, and therefore not shown (see Fig.~\ref{os_g_cuts}). 
  Five million decay events were simulated for each setup.
  \label{ir_g_cuts}}
\end{figure}

The $^{191m}$Ir depopulates via a 2-3 (partly converted) gamma cascade with a total cascade energy of
171~keV. Five million gamma cascades were started inside the detector and the energy
deposition was checked.  Since gamma-rays are much more penetrating than electrons,
the detector exhibited a much more pronounced response.  Fig.~\ref{ir_t_cuts}
shows the response of the array to the gamma-cascade under various thresholds
when only the $^6$LiH shell was in place.  In contrast, Fig.~\ref{ir_g_cuts}
shows the response of the array with no threshold condition on the individual crystal
but with different passive shielding conditions. The detection probability is
calculated as in the case of the osmium decay.  

\begin{figure}
  \includegraphics[width=.75\textwidth]{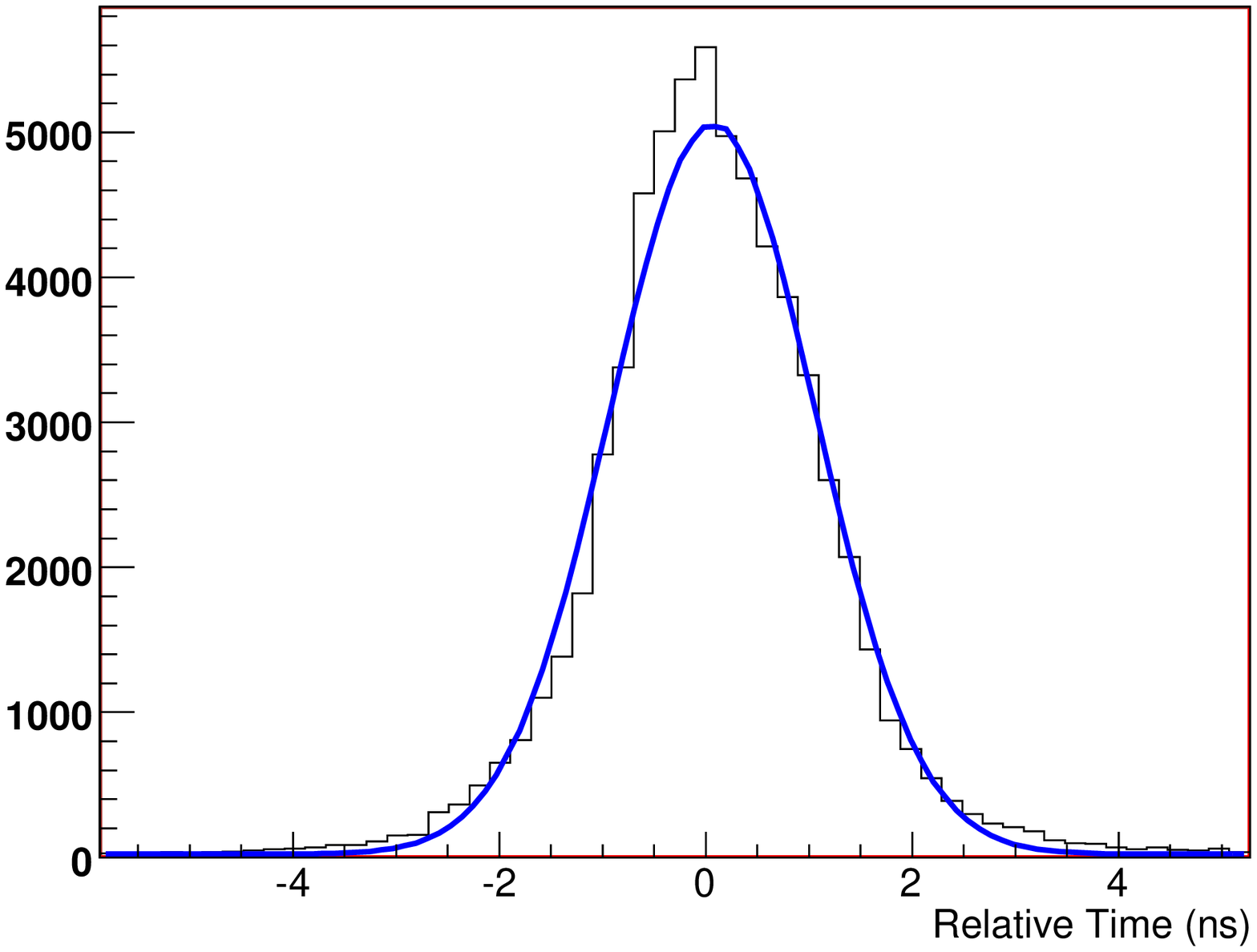}
  \caption{Time distribution of events between two neighboring crystals at 
DANCE on the same digitizer card (\cite{WVB06}).
The histogram shows actual data, while the solid line represents a 
least square fit of a Gaussian curve and a constant background. 
The full width at half maximum (FWHM) of the Gaussian curve is 2.35~ns, 
which means each detector has a time resolution of
  $\pm$0.7~ns. 
  \label{time_resolution}}
\end{figure}

In order to determine the maximum number of atoms of the radioisotope which
can be used in the array, the only thing which remains is to determine the
maximum tolerable rate. Typical BaF$_2$ arrays of that size achieve time resolution of better than 1~ns. 
The Karlsruhe $4\pi$ detector, which has an almost identical geometry as 
the TAC array at n-TOF reports 0.5~ns (\cite{WGK90a}) and
the time resolution measured with the DANCE array was as small as 0.7~ns 
(see Fig.~\ref{time_resolution}). The total width of the coincidence
window for such a detector array can therefore be 5~ns, or even shorter. 
In order to be able to neglect pile-up effects, 
we assumed a maximum event rate of 10$^7$~s$^{-1}$, which means on average 1~event 
every 100~ns. The probability for a 
random coincidence (or a 1-fold pile-up) would then be less than 5\%. 
It should be noted that this is the pile-up rate for the array, not for an
individual crystal.  At this point, 
the reader has to decide what the maximum tolerable rate is under the 
specific conditions of the experiment and scale the numbers found in the table accordingly.   

Then the maximum number of atoms is given by
\begin{equation}
N_{max}\ =\ \frac{10^7}{\lambda (q_{Os}+q_{Ir})} 
       \ =\ 1.92\times10^{13}\, \frac{1}{q_{Os}+q_{Ir}}.
\end{equation}

The average number of detectors firing due to a decay event varies strongly 
depending on the isotope. While it is between one and two for most isotopes, 
it can reach up to six. This is important for estimating the pile-up effect 
within a single detector. An event rate of 10$^7$~s$^{-1}$ (or one event every 100~ns)
would correspond to only 6.25$\cdot$10$^4$~s$^{-1}$ (or one event every 16~$\mu$s) for
multiplicity one events, which is easily manageable. However, it would correspond to a 
detection rate of 3.75$\cdot$10$^5$~s$^{-1}$ (or one event every 2.7~$\mu$s) for multiplicity 
six events. This might require additional measures to avoid pile-up problems. 
The information provided on the maximum sample size for zero threshold can
give some indication as to whether individual detector pile-up will 
introduce significant problems.

The rate of capture events for a particular energy region ($dY/dt$) is given by
$dY/dt\ =\ \sigma\, N_t\, \phi_n$\ where $\sigma$\ is the capture
cross-section over the energy range of interest, $N_t$\ is the number of target
atoms, and $\phi_n$\ is the  neutron flux per second over the same energy region.
The determination of necessary neutron flux was based on the assumption that 
0.1 event per second per energy decade would be necessary for a reasonable 
experiment, so 
\begin{equation}
  \phi_n\ =\ \frac{0.1}{\sigma\, N_{max}}.
\end{equation} 
The cross-section used was the MACS for 30~keV, since this would roughly
correspond to the cross-section for the 10-100~keV decade.

\newpage
\begin{longtable}{c@{\hspace{0.7cm}}ccc|ccc}
\caption{\label{explicit} Extended version of Table~\ref{isotopes} for the example of a $^{191}$Os}\\
\hline
\multicolumn{7}{c}{$^{191}$Os Calculation Table}\\*
&\multicolumn{2}{c}{t$_{1/2}$=15.4} & \multicolumn{2}{c}{$\lambda$=5.21$\times10^{-7}$} &
      \multicolumn{2}{c}{$\sigma$=1290}\\*
&\multicolumn{2}{c}{days} & \multicolumn{2}{c}{decays/sec/atom} & \multicolumn{2}{c}{mb}\\*
\hline
\\*
& \multicolumn{3}{c}{Observed Events} & \multicolumn{3}{c}{Quality factor}\\*
\\*
\hline
$^{191}$Os  & T=0 & T=100 & T=200 & T=0 & T=100 & T=200 \\*
\hline
No Shielding &  356    & 0      &   0   &  7.1$\cdot 10^{-5}$ & 0.0      & 0.0      \\*
LiH Shell &     285    & 1      &   0   &  5.7$\cdot 10^{-5}$ & 2$\cdot 10^{-7}$& 0.0      \\*
LiH+1mm Pb &    0      & 0      &   0   &  0.0      & 0.0      & 0.0      \\*
LiH+2mm Pb &    0      & 0      &   0   &  0.0      & 0.0      & 0.0      \\*
LiH+5mm Pb &    0      & 0      &   0   &  0.0      & 0.0      & 0.0      \\*
\hline
\\
\hline
$^{191m}$Ir  & T=0 & T=100 & T=200 & T=0 & T=100 & T=200 \\ *
\hline
No Shielding & 4580573 & 236879 &   23  &  0.916115 & 0.047376 & 4$\cdot 10^{-6}$ \\*
LiH Shell &    4569545 & 157837 &   16  &  0.913909 & 0.031567 & 3$\cdot 10^{-6}$ \\*
LiH+1mm Pb &   293388  & 8941   &   0   &  0.058678 & 0.001788 & 0.0      \\*
LiH+2mm Pb &   20397   & 499    &   1   &  0.004079 & 1.0$\cdot 10^{-4}$ & 2$\cdot 10^{-7}$\\*
LiH+5mm Pb &   4       & 0      &   0   &  8$\cdot 10^{-7}$& 0.0      & 0.0      \\*
\hline
\\
\hline
Totals & T=0 & T=100 & T=200 & T=0 & T=100 & T=200  \\* 
\hline
No Shielding & 4580929 & 236879 &   23  & 0.916186 & 0.047376  & 4$\cdot 10^{-6}$ \\*
LiH Shell &    4569830 & 157838 &   16  & 0.913966 & 0.031568  & 3$\cdot 10^{-6}$ \\*
LiH+1mm Pb &   293388  & 8941   &   0   & 0.058678 & 0.001788  & 0.0      \\*
LiH+2mm Pb &   20397   & 499    &   0   & 0.004079 & 1.0$\cdot 10^{-4}$  & 2$\cdot 10^{-7}$\\*
LiH+5mm Pb &   4       & 0      &   0   & 8$\cdot 10^{-7}$& 0.0       & 0.0      \\*
\hline
\\
\\
& \multicolumn{3}{c}{Maximum Atoms} & \multicolumn{3}{c}{Required Neutron Flux}\\
& \multicolumn{3}{c}{N$_{atoms}$=$\frac{10^7}{\lambda\cdot q_{tot}}$} & \multicolumn{3}{c}{$\phi_n=\frac{0.1}{\sigma \cdot N_{atoms}}$}\\
\\
\hline
  & T=0 & T=100 & T=200 & T=0 & T=100 & T=200 \\*
\hline
No Shielding &  2.09$\cdot 10^{13}$ &  4.05$\cdot 10^{14}$ &  4.17$\cdot 10^{18}$ & 3.70$\cdot 10^{9}$ &  1.91$\cdot 10^{8}$ & 1.86$\cdot 10^{4}$ \\*
LiH Shell    &  2.10$\cdot 10^{13}$ &  6.08$\cdot 10^{14}$ &  6.00$\cdot 10^{18}$ & 3.69$\cdot 10^{9}$ &  1.27$\cdot 10^{8}$ & 1.29$\cdot 10^{4}$ \\*
LiH+1mm Pb   &  3.27$\cdot 10^{14}$ &  1.07$\cdot 10^{16}$ &  $\infty$            & 2.37$\cdot 10^{8}$ &  7.22$\cdot 10^{6}$ & 0.0 \\*
LiH+2mm Pb   &  4.71$\cdot 10^{15}$ &  1.92$\cdot 10^{17}$ &  9.60$\cdot 10^{19}$ & 1.65$\cdot 10^{7}$ &  4.03$\cdot 10^{5}$ & 8.08$\cdot 10^{2}$ \\*
LiH+5mm Pb   &  2.40$\cdot 10^{19}$ &  $\infty$            &  $\infty$            & 3.23$\cdot 10^{3}$ &  0.0                & 0.0 \\*
\hline
\\
\end{longtable}

\newpage

There are a several of issues to note in Table~\ref{explicit}. First, for brevity, the results
for thresholds with 400 and 600 keV were excluded.  Second, the bottom section differs
slightly from what will be found in the full tables.  In the full tables, if events
were observed in the simulation but the maximum
number of atoms which the system could handle was greater than 10$^{20}$, then no value
was listed as the limitation on the maximum sample size would most likely not
come from the radioactivity of the sample.
Correspondingly, no neutron flux was listed as the neutron flux depends on the number
of target atoms.  These cases are indicated by at ``$*$'' in the final table.  If no
events were observed in the simulation, then there is no limitation on the sample size
from the radioactivity.  Similarly, the necessary neutron flux cannot be calculated.
These cases are indicated by ``---'' in the final table. 

\subsubsection{Scaling Results to Different Instruments}

Before proceeding to present the estimate for sample sizes, it
is appropriate to make a few comments about how one might
estimate the result for a different instrument.  The first 
matter to consider is the maximum tolerable instrument rate.  As was mentioned
before, a rate of 10$^7$~s$^{-1}$\ was assumed for these calculations.  This was
based on the excellent time resolution of BaF$_2$\ detectors and an
observed time coincidence for the DANCE array of better than 5 ns.  The
second consideration is the rate in individual crystals.  Because of the
high segmentation of the DANCE array, the primary limitation came from
the array rate, not the detector rate.  Even in the case of a $\sim$40
element detector, this is likely to be the case.  High segmentation
can alter the observed multiplicity due to increased cross-talk. This
effect which is highly geometry dependent. 
In the case of a very low segmentation instrument ($\sim$10 elements),
the primarily limitation may well come from the individual detector
rate.  Inspection of the decay scheme and benchmarking to
a similar decay would be highly advisable.

\newpage
\noindent 
\begin{longtable}{c@{\hspace{1cm}}ccccc|c}
\caption{\label{isotopes} Master table }\\
\hline
\\*
$^{\mbox{59}}$Fe & \multicolumn{5}{c}{ 18~mb \cite{JEFF30}}\\*
 & \multicolumn{5}{c}{Max. No. of Atoms} &$\phi_n$ \\*
\\*
 & T=0 & T=100 & T=200 & T=400 & T=600 & T=100 \\*
\\*
No Shielding &  6.35$\cdot 10^{13}$ &  6.42$\cdot 10^{13}$ &  6.57$\cdot 10^{13}$ &  7.00$\cdot 10^{13}$ &  7.75$\cdot 10^{13}$ &  8.80$\cdot 10^{10}$ \\
\hline
\end{longtable}

\newpage


\begin{thebibliography}{25}
\expandafter\ifx\csname natexlab\endcsname\relax\def\natexlab#1{#1}\fi
\expandafter\ifx\csname url\endcsname\relax
  \def\url#1{\texttt{#1}}\fi
\expandafter\ifx\csname urlprefix\endcsname\relax\def\urlprefix{URL }\fi

\bibitem[{Apostolakis(1993)}]{GEA93}
Apostolakis, J., 1993. Cern program library long writeup, w5013. Tech. rep.,
  CERN, GEANT library, http://wwwinfo.cern.ch/asd/geant/.

\bibitem[{Bao et~al.(2000)Bao, Beer, K{\"a}ppeler, Voss, Wisshak, and
  Rauscher}]{BBK00}
Bao, Z.~Y., Beer, H., K{\"a}ppeler, F., Voss, F., Wisshak, K., Rauscher, T.,
  2000. Neutron cross sections for nucleosynthesis studies. Atomic Data Nucl.
  Data Tables 76, 70.

\bibitem[{Be{\v{c}}v{\'a}{\v{r}}(2000)}]{bec00}
Be{\v{c}}v{\'a}{\v{r}}, F., 2000. Statistical $\gamma$ cascades following
  thermal and kev neutron capture in heavy nuclei. In: Wender, S. (Ed.),
  Gamma-Ray Spectroscopy and Related Topics. American Institute of Physics, New
  York, p. 504.

\bibitem[{Borcea et~al.(2003)Borcea, Cennini, Dahlfors, Ferrari, Garcia-Munoz,
  Haefner, Herrera-Martinez, Kadi, Lacoste, Radermacher, Saldana, Vlachoudis,
  Zanini, Rubbia, Buono, Dangendorf, Nolte, and Weierganz}]{BCD03}
Borcea, C., Cennini, P., Dahlfors, M., Ferrari, A., Garcia-Munoz, G., Haefner,
  P., Herrera-Martinez, A., Kadi, Y., Lacoste, V., Radermacher, E., Saldana,
  F., Vlachoudis, V., Zanini, L., Rubbia, C., Buono, S., Dangendorf, V., Nolte,
  R., Weierganz, M., Nov 2003. Results from the commissioning of the n-tof
  spallation neutron source at cern. NIM A 513~(3), 524.

\bibitem[{Cano-Ott et~al.(2006)Cano-Ott, Abbondanno, Aerts, Alvarez,
  Alvarez-Velarde, Andriamonje, Andrzejewski, Assimakopoulos, Audouin, Badurek,
  Baumann, Becvar, Benlliure, Berlhoumieux, Calvino, Capote, Albornoz, Cennini,
  Chepel, Chiaveri, Colonna, Cortes, Cortina, Couture, Cox, David, Dolfini,
  Domingo-Pardo, Dridi, Duran, Embid-Segura, Ferrant, Ferrari, Fitzpatrick,
  Ferreira-Marques, Frais-Koelbl, Fujii, Furman, Goncalves, Gallino,
  Gonzalez-Romero, Goverdovski, Gramegna, Griesmayer, Guerrero, Gunsing, Haas,
  Haight, Heil, Herrera-Martinez, Igashira, Isaev, Jericha, Kadi, K{\"a}ppeler,
  Karamanis, Karadimos, Kerveno, Ketlerov, Koehler, Konovalov, Kossionides,
  Krticka, Lamboudis, Leeb, Lindote, Lopes, Lozano, Lukic, Marganiec, Marques,
  Marrone, Mastinu, Mengoni, Milazzo, Moreau, Mosconi, Neves, Oberhummer,
  O'Brien, Oshima, Pancin, Papachristodouiou, Papadopoulos, Papaevangelou,
  Paradela, Patronis, Pavlik, Pavlopoulos, Perdikaki, Perrot, Plag, Plompen,
  Plukis, Poch, Pretel, Quesada, Rauscher, Reifarth, Rosetti, Rubbia, Rudolf,
  Rullhusen, Salgado, Sarchiapone, Stephan, Tagliente, Tain, Tassan-Got,
  Tavora, Terlizzi, Vannini, Vaz, Ventura, Villamarin, Vincente, Vlachoudis,
  Vlastou, Voss, Wendler, Wiescher, and Wisshak}]{CAA06}
Cano-Ott, D., Abbondanno, U., Aerts, G., Alvarez, H., Alvarez-Velarde, F.,
  Andriamonje, S., Andrzejewski, J., Assimakopoulos, P., Audouin, L., Badurek,
  G., Baumann, P., Becvar, F., Benlliure, J., Berlhoumieux, E., Calvino, F.,
  Capote, R., Albornoz, A.~C., Cennini, P., Chepel, V., Chiaveri, E., Colonna,
  N., Cortes, G., Cortina, D., Couture, A., Cox, J., David, S., Dolfini, R.,
  Domingo-Pardo, C., Dridi, W., Duran, I., Embid-Segura, M., Ferrant, L.,
  Ferrari, A., Fitzpatrick, L., Ferreira-Marques, R., Frais-Koelbl, H., Fujii,
  K., Furman, W., Goncalves, I., Gallino, R., Gonzalez-Romero, E., Goverdovski,
  A., Gramegna, F., Griesmayer, E., Guerrero, C., Gunsing, F., Haas, B.,
  Haight, R., Heil, M., Herrera-Martinez, A., Igashira, M., Isaev, S., Jericha,
  E., Kadi, Y., K{\"a}ppeler, F., Karamanis, D., Karadimos, D., Kerveno, M.,
  Ketlerov, V., Koehler, P., Konovalov, V., Kossionides, E., Krticka, M.,
  Lamboudis, C., Leeb, H., Lindote, A., Lopes, I., Lozano, M., Lukic, S.,
  Marganiec, J., Marques, L., Marrone, S., Mastinu, P., Mengoni, A., Milazzo,
  P.~M., Moreau, C., Mosconi, M., Neves, F., Oberhummer, H., O'Brien, S.,
  Oshima, M., Pancin, J., Papachristodouiou, C., Papadopoulos, C.,
  Papaevangelou, T., Paradela, C., Patronis, N., Pavlik, A., Pavlopoulos, P.,
  Perdikaki, G., Perrot, L., Plag, R., Plompen, A., Plukis, A., Poch, A.,
  Pretel, C., Quesada, J., Rauscher, T., Reifarth, R., Rosetti, M., Rubbia, C.,
  Rudolf, G., Rullhusen, P., Salgado, J., Sarchiapone, L., Stephan, C.,
  Tagliente, G., Tain, J.~L., Tassan-Got, L., Tavora, L., Terlizzi, R.,
  Vannini, G., Vaz, P., Ventura, A., Villamarin, D., Vincente, M.~C.,
  Vlachoudis, V., Vlastou, R., Voss, F., Wendler, H., Wiescher, M., Wisshak,
  K., 2006. Neutron capture cross section measurements at n-tof of 237np, 240pu
  and 243am for the transmutation of nuclear waste. AIP Conference Proceedings
  819, 318.

\bibitem[{Firestone(1996)}]{Fir96}
Firestone, R.~B., 1996. Table of Isotopes. Wiley, New York.

\bibitem[{JEFF30 (2005)}]{JEFF30}
JEFF30, 2005. The jeff-3.0 nuclear data library. Tech. rep., JEFF
  Report 19, OECD Nuclear Energy Agency,
  www.nea.fr/html/dbdata/nds\_jefreports/jefreport-19/jefreport-19.pdf.

\bibitem[{K{\"a}ppeler(1999)}]{Kae99}
K{\"a}ppeler, F., 1999. The origin of the heavy elements: the $s$ process.
  Prog. Nucl. Part. Phys. 43, 419 -- 483.

\bibitem[{K{\"a}ppeler et~al.(1989)K{\"a}ppeler, Beer, and Wisshak}]{KBW89}
K{\"a}ppeler, F., Beer, H., Wisshak, K., 1989. s--process nucleosynthesis --
  nuclear physics and the classical model. Rep. Prog. Phys. 52, 945.

\bibitem[{Krane(1988)}]{Kra88}
Krane, K.~S., 1988. Introductory Nuclear Physics. Wiley.

\bibitem[{Lisowski et~al.(1990)Lisowski, Bowman, Russell, and Wender}]{LBR90}
Lisowski, P.~W., Bowman, C.~D., Russell, G.~J., Wender, S.~A., 1990. The los
  alamos national laboratory spallation neutron sources. Nucl. Sci. Engineering
  106, 208.

\bibitem[{Lugaro et~al.(2003)Lugaro, Herwig, Lattanzio, Gallino, and
  Straniero}]{LHL03}
Lugaro, M., Herwig, F., Lattanzio, J.~C., Gallino, R., Straniero, O., 2003.
  s-process nucleosynthesis in asymptotic giant branch stars: A test for
  stellar evolution. Ap.J. 586, 1305.

\bibitem[{Marrone et~al.(2006)Marrone, Abbondanno, Aerts, Alvarez-Velarde,
  Alvarez-Pol, Andriamonje, Andrzejewski, Badurek, Baumann, Becvar, Benlliure,
  Berthomieux, Calvino, Cano-Ott, Capote, Cennini, Chepel, Chiaveri, Colonna,
  Cortes, Cortina, Couture, Cox, Dababneh, Dahlfors, David, Dolfini,
  Domingo-Pardo, Duran-Escribano, Embid-Segura, Ferrant, Ferrari,
  Ferreira-Marques, Frais-Koelbl, Fujii, Furman, Gallino, Goncalves,
  Gonzalez-Romero, Goverdovski, Gramegna, Griesmayer, Gunsing, Haas, Haight,
  Heil, Herrera-Martinez, Isaev, Jericha, Kappeler, Kadi, Karadimos, Kerveno,
  Ketlerov, Koehler, Konovalov, Kritcka, Lamboudis, Leeb, Lindote, Lopes,
  Lozano, Lukic, Marganiec, Martinez-Val, Mastinu, Mengoni, Milazzo,
  Molina-Coballes, Moreau, Mosconi, Neves, Oberhummer, O'Brien, Pancin,
  Papaevangelou, Paradela, Pavlik, Pavlopoulos, Perlado, Perrot, Pignatari,
  Pigni, Plag, Plompen, Plukis, Poch, Policarpo, Pretel, Quesada, Raman, Rapp,
  Rauscher, Reifarth, Rosetti, Rubbia, Rudolf, Rullhusen, Salgado, Soares,
  Stephan, Tagliente, Tain, Tassan-Got, Tavora, Terlizzi, Vannini, Vaz,
  Ventura, Villamarin-Fernandez, Vincente-Vincente, Vlachoudis, Voss, Wendler,
  Wiescher, Wisshak, and n~TOF~Collaborat}]{MAA06}
Marrone, S., Abbondanno, U., Aerts, G., Alvarez-Velarde, F., Alvarez-Pol, H.,
  Andriamonje, S., Andrzejewski, J., Badurek, G., Baumann, P., Becvar, F.,
  Benlliure, J., Berthomieux, E., Calvino, F., Cano-Ott, D., Capote, R.,
  Cennini, P., Chepel, V., Chiaveri, E., Colonna, N., Cortes, G., Cortina, D.,
  Couture, A., Cox, J., Dababneh, S., Dahlfors, M., David, S., Dolfini, R.,
  Domingo-Pardo, C., Duran-Escribano, I., Embid-Segura, M., Ferrant, L.,
  Ferrari, A., Ferreira-Marques, R., Frais-Koelbl, H., Fujii, K., Furman,
  W.~I., Gallino, R., Goncalves, I.~F., Gonzalez-Romero, E., Goverdovski, A.,
  Gramegna, F., Griesmayer, E., Gunsing, F., Haas, B., Haight, R., Heil, M.,
  Herrera-Martinez, A., Isaev, S., Jericha, E., Kappeler, F., Kadi, Y.,
  Karadimos, D., Kerveno, M., Ketlerov, V., Koehler, P.~E., Konovalov, V.,
  Kritcka, M., Lamboudis, C., Leeb, H., Lindote, A., Lopes, M.~I., Lozano, M.,
  Lukic, S., Marganiec, J., Martinez-Val, J., Mastinu, P.~F., Mengoni, A.,
  Milazzo, P.~M., Molina-Coballes, A., Moreau, C., Mosconi, M., Neves, F.,
  Oberhummer, H., O'Brien, S., Pancin, J., Papaevangelou, T., Paradela, C.,
  Pavlik, A., Pavlopoulos, P., Perlado, J.~M., Perrot, L., Pignatari, M.,
  Pigni, M.~T., Plag, R., Plompen, A., Plukis, A., Poch, A., Policarpo, A.,
  Pretel, C., Quesada, J.~M., Raman, S., Rapp, W., Rauscher, T., Reifarth, R.,
  Rosetti, M., Rubbia, C., Rudolf, G., Rullhusen, P., Salgado, J., Soares,
  J.~C., Stephan, C., Tagliente, G., Tain, J.~L., Tassan-Got, L., Tavora, L.
  M.~N., Terlizzi, R., Vannini, G., Vaz, P., Ventura, A., Villamarin-Fernandez,
  D., Vincente-Vincente, M., Vlachoudis, V., Voss, F., Wendler, H., Wiescher,
  M., Wisshak, K., n~TOF~Collaborat, 2006. Measurement of the
  sm-151(n,$\gamma$) cross section from 0.6 ev to 1 mev via the neutron
  time-of-flight technique at the cern n-tof facility. Phys. Rev. C 73, 034604.

\bibitem[{Nassar et~al.(2005)Nassar, Paul, Ahmad, Berkovits, Bettan, Collon,
  Dababneh, Ghelberg, Greene, Heger, Heil, Henderson, Jiang, K{\"a}ppeler,
  Koivisto, O'Brien, Pardo, Patronis, Pennington, Plag, Rehm, Reifarth, Scott,
  Sinha, Tang, and Vondrasek}]{NPA05}
Nassar, H., Paul, M., Ahmad, I., Berkovits, D., Bettan, M., Collon, P.,
  Dababneh, S., Ghelberg, S., Greene, J.~P., Heger, A., Heil, M., Henderson,
  D.~J., Jiang, C.~L., K{\"a}ppeler, F., Koivisto, H., O'Brien, S., Pardo,
  R.~C., Patronis, N., Pennington, T., Plag, R., Rehm, K.~E., Reifarth, R.,
  Scott, R., Sinha, S., Tang, X., Vondrasek, R., 2005. Stellar (n,$\gamma$) cross
  section of ni-62. Phys. Rev. Let. 94, 092504.

\bibitem[{Patronis et~al.(2004)Patronis, Dababneh, Assimakopoulos, Gallino,
  Heil, K{\"a}ppeler, Karamanis, Koehler, Mengoni, and Plag}]{PDA04}
Patronis, N., Dababneh, S., Assimakopoulos, P.~A., Gallino, R., Heil, M.,
  K{\"a}ppeler, F., Karamanis, D., Koehler, P.~E., Mengoni, A., Plag, R., 2004.
  Neutron capture studies on unstable cs-135 for nucleosynthesis and
  transmutation. Phys. Rev. C 69, 025803.

\bibitem[{Rauscher and Thielemann(2001)}]{RaT01}
Rauscher, T., Thielemann, F.-K., 2001. reaction rates. Atomic Data Nucl. Data
  Tables 79, 47.

\bibitem[{Reifarth et~al.(2003{\natexlab{a}})Reifarth, Arlandini, Heil,
  K{\"a}ppeler, Sedychev, Mengoni, Herman, Rauscher, Gallino, and
  Travaglio}]{RAH03}
Reifarth, R., Arlandini, C., Heil, M., K{\"a}ppeler, F., Sedychev, P., Mengoni,
  A., Herman, M., Rauscher, T., Gallino, R., Travaglio, C., 2003{\natexlab{a}}.
  Stellar neutron capture on promethium - implications for the $s$-process
  neutron density. Ap. J. 582, 1251.

\bibitem[{Reifarth et~al.(2004{\natexlab{a}})Reifarth, Bredeweg,
  Alpizar-Vicente, Browne, Esch, Greife, Haight, Hatarik, Kronenberg,
  O'Donnell, Rundberg, Ullmann, Vieira, Wilhelmy, and Wouters}]{RBA04}
Reifarth, R., Bredeweg, T.~A., Alpizar-Vicente, A., Browne, J.~C., Esch, E.-I.,
  Greife, U., Haight, R.~C., Hatarik, R., Kronenberg, A., O'Donnell, J.~M.,
  Rundberg, R.~S., Ullmann, J.~L., Vieira, D.~J., Wilhelmy, J.~B., Wouters,
  J.~M., 2004{\natexlab{a}}. Background identification and suppression for the
  measurement of (n,$\gamma$) reactions with the dance array at lansce. Nucl.
  Instr. Meth. A 531, 528.

\bibitem[{Reifarth et~al.(2003{\natexlab{b}})Reifarth, Haight, Heil, Fowler,
  K{\"a}ppeler, Miller, Rundberg, Ullmann, and Wilhelmy}]{RHH03}
Reifarth, R., Haight, R.~C., Heil, M., Fowler, M.~M., K{\"a}ppeler, F., Miller,
  G.~G., Rundberg, R.~S., Ullmann, J.~L., Wilhelmy, J.~B., 2003{\natexlab{b}}.
  Neutron capture measurements on tm-171. Nucl. Phys. A 718, 478C -- 480C.

\bibitem[{Reifarth et~al.(2004{\natexlab{b}})Reifarth, Haight, Heil,
  K{\"a}ppeler, and Vieira}]{RHH04}
Reifarth, R., Haight, R.~C., Heil, M., K{\"a}ppeler, F., Vieira, D.~J.,
  2004{\natexlab{b}}. Neutron capture measurements at a ria-type facility.
  Nucl. Instr. Meth. A 524, 215.

\bibitem[{Sonnabend et~al.(2003)Sonnabend, Mohr, Vogt, Zilges, Mengoni,
  Rauscher, Beer, K{\"a}ppeler, and Gallino}]{SMV03}
Sonnabend, K., Mohr, P., Vogt, K., Zilges, A., Mengoni, A., Rauscher, T., Beer,
  H., K{\"a}ppeler, F., Gallino, R., 2003. The s-process branching at w-185.
  Ap. J. 583, 506.

\bibitem[{Takahashi and Yokoi(1987)}]{TaY87}
Takahashi, K., Yokoi, K., 1987. Beta-decay rates of highly inonized heavy atoms
  in stellar interiors. Atomic Data Nucl. Data Tables 36, 375.

\bibitem[{Walter et~al.(2006)Walter, Heil, K{\"a}ppeler, Plag, and
  Reifarth}]{WHK06}
Walter, S., Heil, M., K{\"a}ppeler, F., Plag, R., Reifarth, R., 2006. Method
  for (n,g) cross section measurements on unstable isotopes. AIP Conference
  Proceedings 819, 307.

\bibitem[{Wisshak et~al.(1990)Wisshak, Guber, K{\"a}ppeler, Krisch, M{\"u}ller,
  Rupp, and Voss}]{WGK90a}
Wisshak, K., Guber, K., K{\"a}ppeler, F., Krisch, J., M{\"u}ller, H., Rupp, G.,
  Voss, F., 1990. The karlsruhe 4$\pi$ barium fluoride detector. Nucl. Instr.
  Meth. A 292, 595 -- 618.

\bibitem[{Wouters et~al.(2006)Wouters, Vicente, Bredeweg, Esch, Haight,
  Hatarik, O'Donnell, Reifarth, Rundberg, Schwantes, Sheets, Ullmann, Vieira,
  and Wilhelmy}]{WVB06}
Wouters, J.~M., Vicente, A.~A., Bredeweg, T.~A., Esch, E., Haight, R.~C.,
  Hatarik, R., O'Donnell, J.~M., Reifarth, R., Rundberg, R.~S., Schwantes,
  J.~M., Sheets, S.~A., Ullmann, J.~L., Vieira, D.~J., Wilhelmy, J.~B., June
  2006. Acquisition-analysis system for the DANCE (detector for advanced
  neutron capture experiments) BaF$_2$ $\gamma$-ray calorimeter. IEEE
  Transactions on Nuclear Science 53~(3), 880.

\end{thebibliography}

\newcommand{\noopsort}[1]{} \newcommand{\printfirst}[2]{#1}
  \newcommand{\singleletter}[1]{#1} \newcommand{\swithchargs}[2]{#2#1}

\end{document}